# Raman scattering studies of the lateral Mn distribution in MBE-grown $Ga_{1-x}Mn_xN$ epilayers

**Katarzyna Gas[1*], Detlef Hommel[2,3], and Maciej Sawicki[1*]**

[1]*Institute of Physics, Polish Academy of Sciences, Aleja Lotnikow 32/46, PL-02668 Warsaw, Poland*

[2]*Institute of Experimental Physics, University of Wrocław, Pl. Maxa Borna 9, 50-204 Wrocław, Poland*

[3]*Polish Center of Technology Development, ul. Stabłowicka 147, 54-066, Wrocław, Poland*

[*]E-mail: kgas@ifpan.edu.pl, mikes@ifpan.edu.pl

Recent interest in very thin single phase $Ga_{1-x}Mn_xN$ dilute magnetic layers increased needs for precise, non-destructive, and relatively fast characterization methods with key issues being the macroscopic lateral Mn distribution and the absolute values of Mn concentration $x$. We report on resonantly enhanced UV Raman scattering studies of high quality $Ga_{1-x}Mn_xN$ layers grown on GaN templated sapphire by molecular beam epitaxy with $4 < x < 9\%$. The main advantage of the UV excitation is the restriction of the light penetration depth to nearly a hundred nanometers, eliminating signal from the GaN buffer. Under this conditions we determine the dependence of the 1LO phonon frequency on $x$, what allows for a fine mapping of its lateral distribution over the entire surface of the samples. Our Raman scanning clearly confirms substantial lateral distribution of Mn atoms across the layer, which is radial with respect to its center. From the established distributions in two deliberately chosen layers the magnitude of the optimal growth temperature for most efficient Mn atoms incorporation in epitaxial GaN has been confirmed. It is shown that the combination of the 1LO line width and its energy provides assessment of the crystalline quality of the investigated layers.

## 1. Introduction

Ever so increasing importance of ternary and quaternary nitride compounds in fields of blue, ultra-violet (UV), and white light emitting devices [1] and high power applications [2] has increased the requirement on the lateral homogeneity of the relevant layers and quantum structures grown by nonequilibrium growth techniques, the molecular beam epitaxy (MBE) method, in particular. It turns out that apart from such important growth-related parameters like substrate bowing [3] or offcut distribution [4], the lateral inhomogeneity of the growth temperature of the substrate ($T_g$) has equally detrimental influence on the crystalline quality of the epitaxially deposited ternary compounds, as recently evidenced in (Ga,Mn)N [5]. The matter is getting more important not only for the large-size fabrication technology. The concern is also exerted by the basic research community, since, if not recognized prior to processing or functionalization steps, the composition of the investigated device which differs from the assumed one may lead to a considerable misinterpretation of the intended effect or to an undesired functioning of the prospective device. It is therefore of a paramount importance to have identified effective and practical characterization techniques suitable for new materials.

Amongst many non-destructive techniques the Raman spectroscopy offers speed, sensitivity and high lateral resolution, providing that the most convenient excitation range has been applied. This very useful technique allows to characterize semiconductor crystals, thin layers, and low-dimensional structures containing transition metals [6-11] and with a very good spatial resolution [12,13]. In particular, Raman scattering provides information on the crystalline quality, strains, free carrier concentrations (where applicable), and finally on the composition of III-V alloy systems [14-17], disorder [18], as well as it permits large wafer's mapping [19].

(Ga,Mn)N, a magnetic derivative of GaN, in which a fraction $x$ of Ga atoms is randomly replaced by Mn species, has emerged recently as an important for semiconductor spintronics Rashba material [20]. It has also been shown to exhibit promising functionalities after the existence of the direct coupling between the piezoelectricity of wurtzite GaN and the single ion magnetic anisotropy of $Mn^{3+}$ ions [21]. This development has accelerated an interest in very thin, high quality, (Ga,Mn)N magnetic layers, so do the needs for precise and fast characterization. Here, the key issue is the macroscopic lateral distribution of Mn and absolute values of its concentration, as the later has a direct influence on such basic magnetic characteristics as Curie temperature ($T_C$) [5,22,23] and magnetic anisotropy [21]. Since the incorporation of Mn into a semiconductor host during MBE growth process is very sensitive to $T_g$, any lateral inhomogeneity of $T_g$ gives rise to a nonhomogeneous incorporation of Mn, resulting in a strong variations of $T_C$. In a consequence, a negative impact on the functionality of the potential devices is expected, considerably reducing the yield of the fully operational structures. Because of the strong sensitivity of $T_C$ on $x$ ($T_C \sim x^{2.2}$) [22], as in other superexchange-coupled dilute magnetic semiconductors (DMS) [24-27], the lateral distribution of $x$ in homogeneous $Ga_{1-x}Mn_xN$ layers can be quite precisely tracked by magnetic measurements [5]. Both the rather low magnitudes of $T_C$ (below about 13 K) and the highest magnetic saturations among all Mn-containing DMS compounds allow a relatively accurate assessment of $x$, and, as argued recently [28], these features are strongly indicative of a prevalence of Mn in +3 oxidation state. Magnetometry is, generally, a time consuming and, most predominantly, a destructive characterization method, since the original, the as grown material has to be fragmented into small pieces to fit into the sample chamber of the magnetometer. On the other hand, the so much useful high resolution x-ray diffraction (HRXRD) and secondary ion mass spectrometry (SIMS) tools cannot be considered as an alternative high yield every day options, since the former calls for large acquisition times in case of nm thin layers, whereas the latter does destroy the investigated layer, at least at the areas where the depth profiles are taken. Neither the traditional optical methods, as for example the photoluminescence – so indispensable characterization tool of close related nitrides: (In,Ga)N or (Al,Ga)N [29] - can be of a help here. The Fermi level pinning at the mid-gap located $Mn^{3+}$ acceptor in GaN [30,31] effectively kills all the radiative recombination in dilute (Ga,Mn)N. Under such circumstances Raman scattering arises as a very powerful tool, enabling a fast feedback on the relevant material properties in a nondestructive fashion. The required information on the crystalline quality and strains is embedded in the magnitudes of $E_2$ phonon linewidth and frequency, respectively, whereas $A_1$(LO) phonon-plasmon frequencies allow to determine the free carriers concentration. Most importantly here, the frequencies of all phonon modes depend on the alloy composition [32-34]. In particular, for hexagonal $Ga_{1-x}Mn_xN$ with $x \leq 0.2$, except for the $A_1$(TO) phonon, all the other modes are expected to have their frequencies decreasing with the increase of Mn content [35].

To address the need for a fast, accurate, and nondestructive diagnostics for single phase magnetically dilute (Ga,Mn)N layers, we thoroughly examine the lateral dependence of the UV Raman scattering from the LO phonon mode in MBE grown (Ga,Mn)N with Mn concentration ranging from 4.9 to 8.3%. It is shown that combined results of 1LO phonon frequencies and linewidths acquired in a mapping-mode allow to quantify the lateral Mn distribution and provide a qualitative assessment of the layer's structural quality.

## 2. Experiment

About 100 nm thin (Ga,Mn)N layers investigated here have been grown by plasma assisted MBE method in a Scienta-Omicron Pro-100 MBE chamber on GaN(0001) templates deposited on *c*-oriented sapphire in a form of a quarter of the 2 inch wafer. All the pertinent growth conditions and results of the structural characterization are detailed in ref. [5]. The three samples investigated here have been grown at intended $T_g$ of 590, 605, and 620°C, while keeping all the other growth parameters nominally unchanged. Accordingly, the samples are labeled as S590, S605, and S620. These temperatures have been set according to the read-out of a video pyrometer aimed at the *center* of the substrate. For these magnitudes of $T_g$ the Mn incorporation in the central parts of the samples amounts to about 7, 8, and 6%, respectively, as determined jointly by magnetic saturation at low temperatures, SIMS, and HRXRD [5]. However, due to the existence of the distribution of $T_g$ across the substrates and a rather strong dependence of *x* on $T_g$, the overall distribution of *x* in these three layers extends from about 5 to 9.5%, as assessed previously in ref. [5], and detailed later in this letter. The existence of the optimum value of $T_g$ for achieving the most efficient Mn incorporation in GaN at the given growth conditions employed by us (about 605 +/- 5 °C) results in a situation that the central parts of samples S620 and S605 exhibit the highest, whereas the centre of S590 sample exhibits the lowest *x* found in each of these samples. The detailed in-depth magnetic studies, performed according to the lab-code [36] and tools [37] elaborated for studies of minute magnetic moments deposited on bulky substrates, established *the single magnetic phase* constitution of the studied layers [5].

The room temperature micro-Raman scattering measurements are performed using a Monovista CRS+ Confocal Laser Raman System with the spectral resolution below 1 cm$^{-1}$. The spectra are excited with two laser wavelengths $\lambda_{exc}$ = 532 and 325 nm. After focusing on the specimen the spot diameters of the incident light amount to 1 and 10 μm, respectively. The measurements are performed in the backscattering geometry. According to the selection rules for experimental geometry employed here three phonons E$_2$(low), E$_2$(high) and A$_1$(LO) in the first order Raman spectrum of GaN are allowed [7][16]. Overtones always contain the representation A$_1$ while combinations of phonon states belonging to different irreducible representations never contain A$_1$ representation [38]. Care has been taken to ensure that the illumination does not lead to any heat developments in the investigated films. The polarization of the scattered light has not been analyzed. The position of the Raman peaks as well as their full width at half maximum (FWHM) values are obtained from a fitting of the Lorentzian shape to the relevant sections of the spectra with the typical deviations not exceeding 0.3 and 1 cm$^{-1}$, respectively.

## 3. Results and discussion

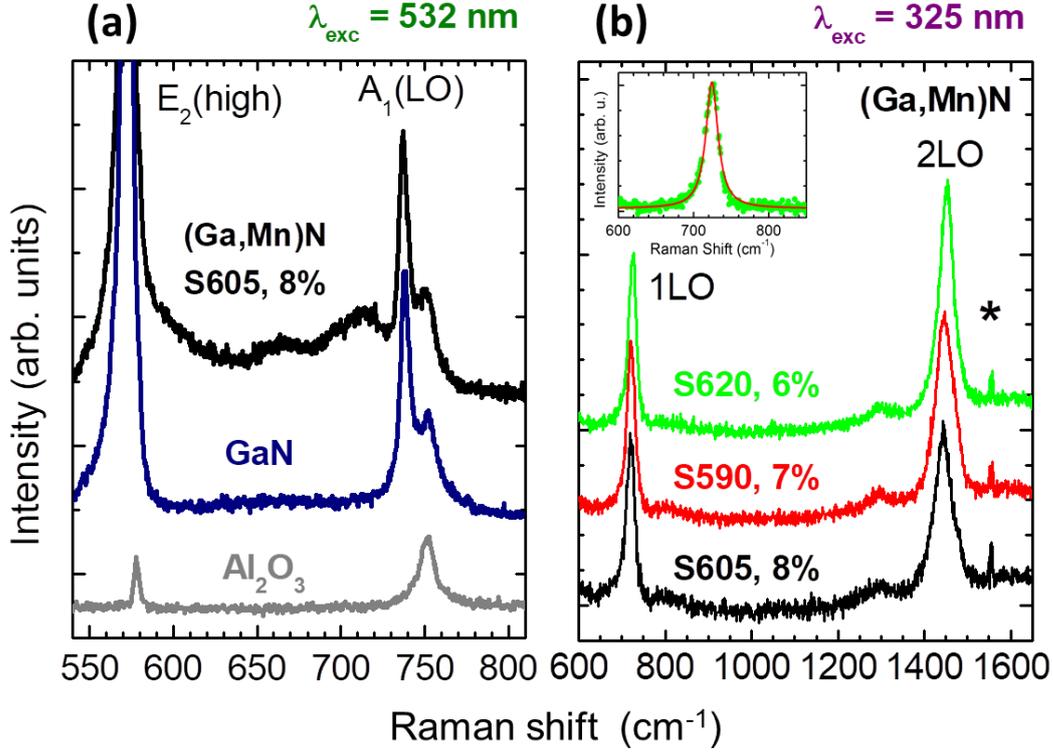

**Fig. 1.** a) Exemplary Raman spectra obtained with $\lambda_{exc}$ = 532 nm laser excitation. (From top to bottom): (Ga,Mn)N epilayer (sample S605, black) and two references: GaN template on $Al_2O_3$ (navy) and the bare $Al_2O_3$ substrate (grey). The spectra of both (Ga,Mn)N and GaN epilayers are normalized by the peak intensity of the most intensive $E_2$(high) GaN phonon mode, whereas the spectrum of $Al_2O_3$ has been scaled to reproduce the height of the $E_g$ $Al_2O_3$ mode, seen at ~750 cm$^{-1}$ in all three spectra. b) 1LO and 2LO-phonon peaks observed under the above band-gap excitation (325 nm laser light) of (Ga,Mn)N layers with different Mn concentrations established at the central part of the as grown sample. The spectra are normalized by the 1LO phonon peak intensity and shifted vertically for clarity. Features marked by asterisk are ambient $O_2$ bands. The spectra are shifted vertically for clarity of the presentation. Inset: a zoom of the 1LO peak region of the S620 sample with the data points in green and the Lorenzian fit (the red line).

The representative high frequency part of the Raman spectrum of one of the (Ga,Mn)N layer (sample S605, $x \cong 8\%$) excited with laser wavelength of 532 nm is compared with the GaN template and an $Al_2O_3$ substrate spectra in Fig. 1 (a). The data exemplifies the fact that Raman scattering in visible frequency range does not provide relevant information on thin (Ga,Mn)N layers. The absorption coefficient for this wavelength in GaN is negligibly small [39], so the light penetrates through both the whole (Ga,Mn)N and GaN layers into the $Al_2O_3$ substrate. For this reason the sample's Raman spectrum contains mostly the signal from the 3 µm thick GaN buffer, rendering the analysis of the phonon modes in ~100 nm thin (Ga,Mn)N layer rather problematic. Indeed, the dominant features in the (Ga,Mn)N spectrum, the $E_2$(high) and $A_1$(LO) peaks at approximately at 570 and 735 cm$^{-1}$, respectively, are found at exactly the same position as in the GaN template. Both spectra exhibit also peaks of $Al_2O_3$ substrate at about 578 and 751 cm$^{-1}$. The main difference between (Ga,Mn)N and GaN buffer spectra is seen around 600, 665, and 710 cm$^{-1}$, where weak and broad bands develop in (Ga,Mn)N. These phonon structures ubiquitously observed in (Ga,Mn)N mixed crystals [7,35,40-43] are understood to reflect the vibration modes of the distorted nitrogen sublattice [7]. However, they are too weak and too broad to serve as a reliable information source on the constitution of the investigated layer.

In order to restrict the probed depth of the samples to just a few tens of nanometers we take advantage of the much shallower penetration of the 325 nm (3.82 eV) UV light into the material, which according to ref. 39 should be less than 100 nm for all values of $x$ studied here. This ensures that most of the Raman-scattering signal comes from the (Ga,Mn)N layers, thus eliminating the detrimental contributions of the luminescence from the GaN buffer layer. The energy of the excitation photons is close to the band-gap of GaN (3.4 eV at room temperature) and (Ga,Mn)N, since the band-gap opening rate in (Ga,Mn)N amounts only to about 0.027 eV per % of Mn [44]. Accordingly, even with the highest Mn content of about 9.5% the band gap of (Ga,Mn)N layer does not exceed 3.7 eV, thus for all measured samples the conditions for the above band-gap excitation are satisfied.

The representative Raman spectra obtained under 325 nm excitation from the central parts of the three (Ga,Mn)N epilayers studied here are shown in Fig. 1 (b). For this wavelength due to Fröhlich interaction [45][46] strong resonant Raman scattering of the A$_1$(LO) phonon relative to E$_2$(high) as well as the second and the third-order (not shown) Raman scattering is observed. The frequency of the 1LO phonon mode ($\omega_{LO}$) changes form (724.2 ± 0.1) cm$^{-1}$ for the sample S620 ($x \cong 6\%$) to (719.5 ± 0.2) cm$^{-1}$ for the sample S605 ($x \cong 8\%$). A representative Lorentzian fit to the experimental data is presented in the inset to Fig. 1 (b). Such a beyond the experimental uncertainty dependence of $\omega_{LO}$ on $x$, indicates that UV Raman scattering can be employed to relatively accurate determination the Mn content in (Ga,Mn)N compound. However, to obtain absolute magnitudes of $x$ a proper calibration of this tool has to be performed first.

To this end we consider two samples, S620 and S605, in which (Ga,Mn)N layers had been found fully strained with respect to the GaN buffer, assuring that their lattice parameters $a$ and $c$ remain in the same epitaxial relationship with $x$, as found in ref. [5]. The same studies showed that the sample grown at the lowest $T_g$ (S590) exhibits partial relaxation of the lattice. Since the magnitude of $c$, and so $\omega_{LO}$ in this layer, may not reflect $x$ in the same degree as it does in the fully strained layers, this sample is excluded from our consideration at this stage. Nevertheless, the two first layers provide a sufficiently wide range of $x$ to assess the character of $\omega_{LO}(x)$. This is because of the nonuniform heat transfer to the sapphire substrates in the MBE growth chamber leading to sizable lateral variation of $T_g$, reaching $\Delta T_g \cong 12$ °C between the edge and the center of approximately 1" wide substrate. For example, such magnitude of $\Delta T_g$ yields $\Delta x$ as high as ~2% in 7% S620 sample [5].

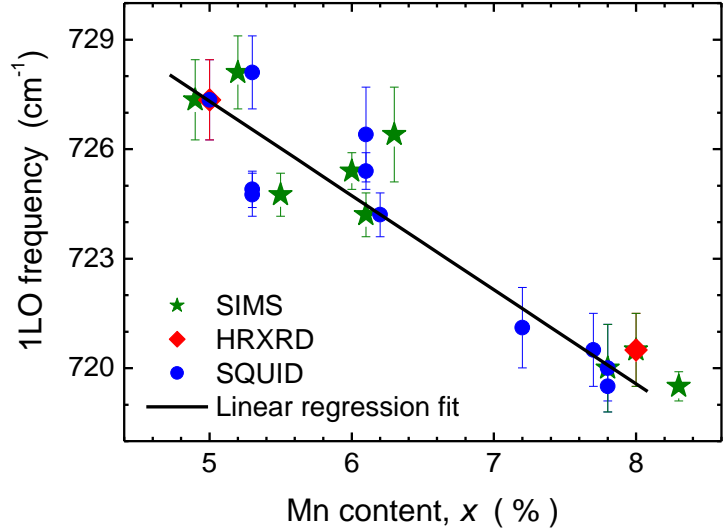

**Fig. 2.** 1LO phonon frequency plotted versus Mn content ($x$) determined by secondary ion mass spectrometry (SIMS, green stars), high resolution X-ray diffraction (HRXRD, red diamonds), and from low temperature saturation magnetization (SQUID, blue bullets) [5]. Solid line is the linear regression fit to the data given in Eq. 1.

In Fig. 2 we plot the magnitudes of $\omega_{LO}$ established for a range of specimens cleaved from samples S620 and S605 as a function of $x$, determined for these fragments independently by HRXRD, SIMS and/or SQUID magnetometry. The sizes of investigated specimens are between 12 and 25 mm$^2$. Each value of $\omega_{LO}$ indicated in the Fig. 2 represents the average of 4 – 6 readings taken across each investigated fragment. This is to simulate the determination of $x$ from HRXRD and SQUID measurements, which are integral probes, so their outcome represents an averaged magnitude of $x$ in the given specimen. The magnitudes of the error bars for the phonon line positions are given by the standard deviation of these averages. The results presented in Fig. 2 tend to group in pairs, or even triplets, since all the available evaluations of $x$ are indicated in the figure. As seen in the figure, a set of specimens cut from these nominally $x$ = 6 and 8% (Ga,Mn)N layers covers the range of $x$ from 4.9 to 8.3%. The main feature provided by this plot is that $\omega_{LO}$ decreases with $x$, a fact anticipated on the account of the increase of the $c$ lattice parameter on $x$ in (Ga,Mn)N epilayers [5][23]. This assumption is fully corroborated by the insulating character of magnetically homogeneous (Ga,Mn)N bulk crystals [47] and epilayers [28,48-50], thus no influence of phonon-plasmon coupling is expected here.

Now, guided by the spread of the points in Fig. 2, we assume the simplest, a linear dependency $\omega_{LO}(x)$, which upon the linear regression decreases (in the units of Fig. 2) as:

$$\omega_{LO} = (739.4 \pm 1.2) - (2.4 \pm 0.2)\, x. \qquad (1)$$

It has to be noted here, that this linear approximation is valid only for these rather large Mn concentrations. For $x \to 0$ the dependency given by Eq. 1 overshoots the A$_1$(LO) energy in GaN by about 4.5 cm$^{-1}$. This fact indicates a more complex form of $\omega_{LO}(x)$ in the whole range of $x$, an effect previously observed [51,52], and expected theoretically by first principle calculations [35]. The limited accuracy of the linear approximation employed here is also evident from the visible spread of the experimental points, and confirmed by the correlation coefficient $r = -0.94$. Nevertheless, the relative small errors of the linear and angular coefficients allow us to perform relatively fast and nondestructive diagnostics of the lateral distribution of $x$. Interestingly, the slope $s \cong 2.4$ cm$^{-1}$ per percent of Mn lies between previously found values of $s \cong 0.8$ and 5.6 cm$^{-1}$ per percent of Mn in Mn-implanted GaN [51] and in nanocrystalline films deposited by reactive magnetron sputtering [52], respectively. Such a discrepancy indicates rather sizable structural differences between (Ga,Mn)N obtained by these two approaches and the MBE technique. From this perspective we underline that the $\omega_{LO}$ dependency on $x$ in Ga$_{1-x}$Mn$_x$N has been established for the first time for single phase epitaxial layers with such high magnitudes of $x$. Finally, and importantly for the present study, at the investigated here range of $x > 4.8\%$, the magnitude of $s$ given by Eq. 1 allows a fine $x$-mapping of the material.

Having our Raman spectroscopy tool calibrated we can demonstrate that the monitoring of the magnitude of $\omega_{LO}$ can be used to determine the extent and the form of the lateral distribution of Mn concentration over the (Ga,Mn)N surfaces. This is illustrated in Fig. 3, where both the frequencies of the 1LO phonon modes (bottom panels) and their FWHM (top panels) are plotted versus the positions from which they were acquired on the substrate. The two samples presented in the figure, S605 and S590, are scanned in a linear fashion, with the scanning line passing in parallel to one of the substrate's straight sides and about 1.6 cm away from it. The clearly higher wavenumbers of the LO modes measured at the both ends of the scan of sample S605, panel (c), are indicative of a lower Mn concentration at the edges in comparison with the center of this sample. The magnitudes of $x$ established upon Eq. 1 are indicated on the right Y axis of Fig. 3(c), yielding a change $|\Delta x| \cong 2.8\%$, from ~8% near the center to ~5.2% at the edge. This is a particularly dramatic change for DMS, and generally for dilute magnetic systems, since their characteristic magnetic temperatures depend either linearly on $x$, as $T_C$ in (Ga,Mn)As [53], or quadratically in such short range coupled systems [22,24,54], leading to a rather large 40% variations of $T_C$ in (Ga,Mn)N [5]. This negative gradient of $x$ (counting from the center) is confirmed by the opposite trend of FWHM, depicted in panel (a), since the FWHM is expected to grow on $x$ [51,52], so it peaks at the center area. In the case of sample S590 the scanning yields a reversed dependence of $\omega_{LO}$ on position. For this low $T_g$ the Mn concentration is the lowest at the coldest central part of the substrate and increases towards the warmer edges. This observation yet again confirms the existence of the optimal growth temperature above and below which the incorporation of Mn decreases.

On the other hand, it is worth noting that the absolute values of 1LO frequencies differ for these two samples even in locations where the same Mn content is expected according to before-growth $T_g$ assessment performed by the pyrometer. This finding underlines the remarkable sensitivity of the Raman scattering to even small structural imperfections of the investigated material. The leading effect in this case is a partial strain relaxation in layer S590, whereas layer S605 retains the perfect pseudomorphic relation to the buffer. Despite this difference, the clearly concave shape seen in Fig. 3(d) indicates that very detailed qualitative information on Mn distribution across the layer can still be extracted. Nevertheless, to indicate their only indicatory character, the magnitudes of $x$ established for sample S590 are marked in panel (d) in a gray shade. Remarkably, this slightly lower crystalline quality is perfectly reflected by a featureless character of FWHM dependence on position, presented in Fig. 3(b). Such a lack of accord between the results presented in panels (b) and (d) does confirm a mediocre crystalline quality of this partially relaxed layer. On the other hand, a very positive outcome of this study is the finding that both $\omega_{LO}$ and FWHM are constant within about 8 mm wide region of the central part of the S605 layer, indicating the existence of a regions with fairly uniform concentration of Mn.

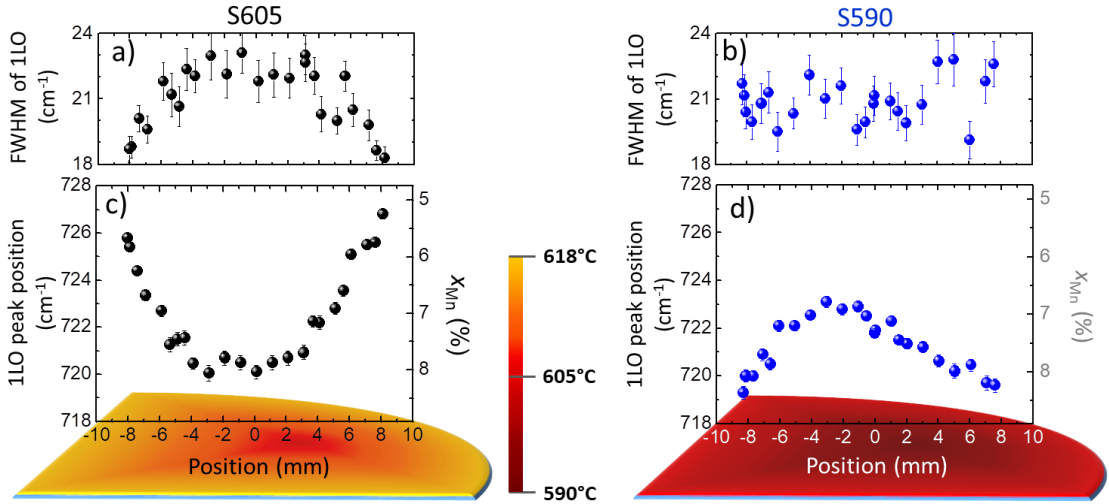

**Fig. 3**. Lateral dependence of the frequencies of the 1LO phonon modes (bottom panels) and their FWHM (top panels) for samples S605 (left panels) and S590 (right panels). The linear scans are taken in parallel to and about 1.6 cm away from one of the straight edges of the quarter of a 2" wafer substrate, as indicated at the bottom drawings. The distribution of the growth temperature across the substrates is rendered on these drawings following live temperature readings of a video pyrometer. Right Y-axes of the bottom panels indicate Mn concentration obtained upon the calibration given by Eq. 1. Nevertheless, to indicate their only indicatory character, the magnitudes of $x$ established for sample S590 are marked in panel (d) in a gray shade.

### 4. Conclusions

Resonantly enhanced UV Raman spectroscopy has been used to develop a fast and precise tool to determine the Mn content and assess the crystalline quality of ternary (Ga,Mn)N insulating thin epitaxial layers in a non-destructive manner. This is possible here only due to the Mn-related quenching of the luminescence, which otherwise obscures phonon-related features for UV-excitation in GaN-based compounds. The quantitative results of the calibration presented here are valid for fully strained layers and allow for immediate application of the method in the concentration range between 4 and 9%, i.e. where the truly magnetically dilute layers exhibit ferromagnetic properties at low end of the cryogenic temperatures. The implementation of the method for fast concentration assessments in samples obtained by other growth techniques and in suitable material systems is just straightforward, but for reliable quantitative information an own calibration effort has to be exerted. In our particular case, the Raman spectroscopy has allowed to precisely establish the optimum growth temperature of (Ga,Mn)N by a relatively accurate scanning across the surface of layers grown with a non-uniformly heated substrates, substantiating previous studies on this subject. By the same token regions of the most uniform or simply desired Mn concentration can be easily identified. In a broader context, the method presented here, by its ease, simplicity and malleability for automatization can sizably aid the technological efforts to elaborate recipes for increase the concentration of substitutional Mn cations in GaN beyond 20%, so a vital achievement in order to bring up the ferromagnetic phase transition in this technologically profoundly important family of nitrides to the vicinity of the room temperature.

**Acknowledgments**

This work has been supported by the National Science Centre, Poland through OPUS (DEC-2018/31/B/ST3/03438) call grant.

**References**

[1] H. Morkoc, Handbook of Nitride Semiconductors and Devices: Electronic and Optical Processes in Nitrides, Wiley-VCH Verlag GmbH, Weinheim 63 (1) (2016) 419–425.

[2] H. Amano, Y. Baines, E. Beam, M. Borga, T. Bouchet, P.R. Chalker, M. Charles, K.J. Chen, N. Chowdhury, R. Chu, C. De Santi, M.M. De Souza, S. Decoutere, L. Di Cioccio, B. Eckardt, T. Egawa, P. Fay, J.J. Freedsman, L. Guido, O. Häberlen, G. Haynes, T. Heckel, D. Hemakumara, P. Houston, J. Hu, M. Hua, Q. Huang, A. Huang, S. Jiang, H. Kawai, D. Kinzer, M. Kuball, A. Kumar, K.B. Lee, X. Li, D. Marcon, M. März, R. McCarthy, G. Meneghesso, M. Meneghini, E. Morvan, A. Nakajima, E.M.S. Narayanan, S. Oliver, T. Palacios, D. Piedra, M. Plissonnier, R. Reddy, M. Sun, I. Thayne, A. Torres, N. Trivellin, V. Unni, M.J. Uren, M. Van Hove, D.J. Wallis, J. Wang, J. Xie, S. Yagi, S. Yang, C. Youtsey, R. Yu, E. Zanoni, S. Zeltner, Y. Zhang, The 2018 GaN power electronics roadmap, J. Phys. D: Appl. Phys. 51 (2018) 163001.

[3] T. Sochacki, Z. Bryan, M. Amilusik, R. Collazo, B. Lucznik, J. L. Weyher, G. Nowak, B. Sadovyi, G. Kamler, R. Kucharski, M. Zajac, R. Doradzinski, R. Dwilinski, I. Grzegory, M. Bockowski, Z. Sitar, Preparation of Free-Standing GaN Substrates from Thick GaN Layers Crystallized by Hydride Vapor Phase Epitaxy on Ammonothermally Grown GaN Seeds, Appl. Phys. Express 6 (2013) 075504.

[4] I. Bryan, Z. Bryan, S. Mita, A. Rice, L. Hussey, Ch. Shelton, J. Tweedie, J.P. Maria, R. Collazo, Z. Sitar, The role of surface kinetics on composition and quality of AlGaN, Journal of Crystal Growth 451, (2016) 65.

[5] K. Gas, J.Z. Domagala, R. Jakiela, G. Kunert, P. Dluzewski, E. Piskorska-Hommel, W. Paszkowicz, D. Sztenkiel, M.J. Winiarski, D. Kowalska, R. Szukiewicz, T. Baraniecki, A. Miszczuk, D. Hommel, M. Sawicki, Impact of substrate temperature on magnetic properties of plasma-assisted molecular beam epitaxy grown (Ga,Mn)N, J. Alloys Compd. 747 (2018) 946.

[6] H. Harima, Raman studies on spintronics materials based on wide bandgap semiconductors, J. Phys. Condens. Matter. 16, (2004) S5653–S5660.

[7] W. Gebicki, P. Dominik, S. Podsiadlo, Lattice dynamics and Raman scattering from GaN:Mn crystals, Phys. Rev. B 77 (2008) 245213.

[8] W. Szuszkiewicz, M. Jouanne, J.F. Morhange, M. Kanehisa, E. Dynowska, K. Gas, E. Janik, G. Karczewsk, R. Kuna, J. Sadowski, T. Wojtowicz, Raman scattering as a tool to characterize semiconductor crystals, thin layers, and low-dimensional structures containing transition metals, Phys. Status Solidi B 251 (2014) 1133.

[9] K. Gas, E. Janik, W. Zaleszczyk, E. Dynowska, M. Kutrowski, A. Kamińska, J.F. Morhange, Ł. Wachnicki, T. Wojciechowski, R. Hołyst, M. Godlewski, E. Guziewicz, T. Wojtowicz, W. Szuszkiewicz, Selected optical properties of core/shell ZnMnTe/ZnO nanowire structures, Phys.

Status Solidi B 248 (2011) 1592.

[10] K. Gas, J. Sadowski, J.F. Morhange, A. Siusys, W. Zaleszczyk, T. Wojciechowski, T. Kasama, A. Altintaş, H.Q. Xu, W. Szuszkiewicz, Structural and optical properties of self-catalytic GaAs:Mn nanowires grown by molecular beam epitaxy on silicon substrates, Nanoscale 5 (2013) 7410.

[11] A. Siusys, J. Sadowski, M. Sawicki, S. Kret S, T. Wojciechowski, K. Gas, W. Szuszkiewicz, A. Kamińska, T. Story, All-Wurtzite (In,Ga)As-(Ga,Mn)As Core Shell Nanowires Grown by Molecular Beam Epitaxy, Nano Lett. 14 (2014) 4263.

[12] Y. Huang, X.D. Chen, S. Fung, C.D. Beling, C.C. Ling, Spatial characterization of a 2 in GaN wafer by Raman spectroscopy and capacitance–voltage measurements, J. Phys. D: Appl. Phys. 37 (2004) 2814.

[13] M. Amilusik, D. Wlodarczyk, A. Suchocki, M. Bockowski, Micro-Raman studies of strain in bulk GaN crystals grown by hydride vapor phaseepitaxy on ammonothermal GaN seeds, Jpn. J. Appl. Phys.58 (2019) SCCB32.

[14] L. Artús, R.A. Stradling, Y.B. Li, S.J. Webb, W.T. Yuen, S.J. Chung, R. Cuscó, Raman scattering from optical phonons in $InAs_{1-x}Sb_x$/InAs strained-layer superlattices, Phys. Rev. B 54 (1996) 16373.

[15] M. Kuball, J.M. Hayes, T. Suski, J. Jun, M. Leszczynski, J. Domagala, H.H. Tan, J.S. Williams, C. Jagadish, High-pressure high-temperature annealing of ion-implanted GaN films monitored by visible and ultraviolet micro-Raman scattering, J. Appl. Phys. 87 (2000) 2736.

[16] H. Harima, Properties of GaN and related compounds studied by means of Raman scattering, Journal of Physics: Condensed Matter 14 (2002) R967.

[17] W. Ager III, W. Walukiewicz, W. Shan, K. M. Yu, S. X. Li, E. E. Haller, H. Lu, W. J. Schaff, Multiphonon resonance Raman scattering in $In_xGa_{1-x}N$, Phys. Rev. B 72 (2005) 155204.

[18] H. J. Trodahl, F. Budde, B. J. Ruck, S. Granville, A. Koo, A.Bittar, Raman spectroscopy of nanocrystalline and amorphous GaN, J. Appl. Phys. 97 (2005) 084309.

[19] M. M. Muhammed, M. A. Roldan, Y. Yamashita, S. L. Sahonta, I. A. Ajia, K. Iizuka, A. Kuramata, C. J. Humphreys, I. S. Roqan, High-quality III-nitride films on conductive, transparent (201)-oriented β-$Ga_2O_3$ using a GaN buffer layer, Sci. Rep. 6 (2016) 29747.

[20] W. Stefanowicz, R. Adhikari, T. Andrearczyk, B. Faina, M. Sawicki, J. A. Majewski, T. Dietl, A. Bonanni, Experimental determination of Rashba spin-orbit coupling in wurtzite n-GaN:Si, Phys. Rev. B 89 (2014) 205201.

[21] D. Sztenkiel, M. Foltyn, G. P. Mazur, R. Adhikari, K. Kosiel, K. Gas, M. Zgirski, R. Kruszka, R. Jakiela, Tian Li, A. Piotrowska, A. Bonanni, M. Sawicki, T. Dietl, Stretching magnetism with an electric field in a nitride semiconductor, Nature Commun. 7 (2016) 13232.

[22] S. Stefanowicz, G. Kunert, C. Simserides, J.A. Majewski, W. Stefanowicz, C. Kruse, S. Figge, T. Li, R. Jakieła, K.N. Trohidou, A. Bonanni, D. Hommel, M. Sawicki, T. Dietl, Phase diagram and critical behavior of the random ferromagnet $Ga_{1-x}Mn_xN$, Phys. Rev. B. 88 (2013) 081201.

[23] G. Kunert, S. Dobkowska, T. Li, H. Reuther, C. Kruse, S. Figge, R. Jakieła, A. Bonanni, J. Grenzer, W. Stefanowicz, J. von Borany, M. Sawicki, T. Dietl, D. Hommel, $Ga_{1-x}Mn_xN$ epitaxial films with high magnetization, Appl. Phys. Lett. 101 (2012) 022413.

[24] M. Sawicki, E. Guziewicz, M. I. Lukasiewicz, O. Proselkov, I.A. Kowalik, W. Lisowski, P. Dłużewski, A. Wittlin, M. Jaworski, A. Wolska, W. Paszkowicz, R. Jakieła, B. S. Witkowski, L. Wachnicki, M. T. Klepka, F. J. Luque, D. Arvanitis, J. W. Sobczak, M. Krawczyk, A. Jablonski, W. Stefanowicz, D. Sztenkiel, M. Godlewski, T. Dietl, Homogeneous and heterogeneous magnetism in (Zn,Co)O: From a random antiferromagnet to a dipolar superferromagnet by changing the growth temperature, Phys. Rev. B 88 (2013) 085204.

[25] P. W. Anderson, Antiferromagnetism. Theory of Superexchange Interaction, Phys. Rev. 79 (1950) 350.

[26] J. B. Goodenough, An interpretation of the magnetic properties of the perovskite-type mixed crystals

La$_{1-x}$Sr$_x$CoO$_{3-\lambda}$, J. Phys. Chem. Solids 6 (1958) 287.

[27] J. Kanamori, Superexchange interaction and symmetry properties of electron orbitals, J. Phys. Chem. Solids 10 (1959) 87.

[28] K. Kalbarczyk, K. Dybko, K. Gas, D. Sztenkiel, M. Foltyn, M. Majewicz, P. Nowicki, E. Łusakowska, D. Hommel, M. Sawicki, Electrical characteristics of vertical-geometry Schottky junction to magnetic insulator (Ga,Mn)N heteroepitaxially grown on sapphire, J. Alloys Compd. 804 (2019) 415-420.

[29] M. Siekacz, A. Feduniewicz-Zmuda, G. Cywiński, M. Kryśko, I. Grzegory, S. Krukowski, K.E. Waldrip, W. Jantsch, Z.R. Wasilewski, S. Porowski, C. Skierbiszewski, Growth of InGaN and InGaN/InGaN quantum wells by plasma-assisted molecular beam Epitaxy, J. Cryst. Growth. 310 (2008) 3983-3986.

[30] A. Wolos, M. Palczewska, M. Zajac, J. Gosk, M. Kaminska, A. Twardowski, M. Bockowski, I. Grzegory, S. Porowski, Optical and magnetic properties of Mn in bulk GaN, Phys. Rev. B 69 (2004) 115210.

[31] L. Janicki, G. Kunert, M. Sawicki, E. Piskorska-Hommel, K. Gas, R. Jakiela, D. Hommel, R. Kudrawiec, Fermi level and bands offsets determination in insulating (Ga,Mn)N/GaN structures, Scientific Reports 7 (2017) 41877.

[32] H. Grille, Ch. Schnittler, F. Bechstedt, Phonons in ternary group-III nitride alloys, Phys. Rev. B 61 (2000) 6091.

[33] S. Hernández, R. Cuscó, D. Pastor, L. Artús, K. P. O'Donnell, R. W. Martin, I. M. Watson, Y. Nanishi and E. Calleja, Raman-scattering study of the InGaN alloy over the whole composition range, J. Appl. Phys. 98 (2005) 013511.

[34] R. Oliva, J. Ibanez, R. Cusco, R. Kudrawiec, J. Serafinczuk, O. Martinez, J. Jimenez, M. Henini, C. Boney, A. Bensaoula, L. Artus, Raman scattering by the E$_{2h}$ and A$_1$(LO) phonons of In$_x$Ga$_{1-x}$N epilayers (0.25 < $x$ < 0.75) grown by molecular beam epitaxy, J. Appl. Phys. 111 (2012) 063502.

[35] H. W. Leite Alves, L. M. R Scolfaro, E. F. da Silva Jr, Lattice dynamics of Ga$_{1-x}$Mn$_x$N and Ga$_{1-x}$Mn$_x$As by first-principle calculations, Nanoscale Research Letters 7 (2012) 573.

[36] M. Sawicki, W. Stefanowicz, A. Ney, Sensitive SQUID magnetometry for studying nanomagnetism, Semicond. Sci. Technol. 26 (2011) 064006.

[37] K. Gas and M. Sawicki, In situ compensation method for high-precision and high-sensitivity integral magnetometry, Meas. Sci. Technol. 30 (2019) 085003.

[38] H. Siegle, G. Kaczmarczyk, L. Filippidis, A. P. Litvinchuk, A. Hoffmann, and C. Thomsen, Zone-boundary phonons in hexagonal and cubic GaN, Phys. Rev. B 55 (1997) 7000.

[39] J. F. Muth, J. H. Lee, I. K. Shmagin, R. M. Kolbas, H. C. Casey, Jr, B. P. Keller, U. K. Mishra, S. P. DenBaars, Absorption coefficient, energy gap, exciton binding energy, and recombination lifetime of GaN obtained from transmission measurements, Appl. Phys. Lett. 71 (1997) 2572.

[40] X. Yang, J. Wu, Z. Chen, Y. Pan, Y. Zhang, Z. Yang, T. Yu, G. Zhang, Raman scattering and ferromagnetism of (Ga, Mn)N films grown by MOCVD, Solid State Commun. 143 (2007) 236.

[41] N. Hasuike, H. Fukumura, H. Harima, K. Kisoda, M. Hashimoto, Y.K. Zhou, H. Asahi, Optical studies on GaN-based spintronics materials, J. Phys.: Condens. Matter, 16 (2004) S5811.

[42] M. Zajac, J. Gosk, M. Kaminska, A. Twardowski, T. Szyszko, S. Podsiadło, Magnetic and optical properties of GaMnN magnetic semiconductor, Appl. Phys. Lett. 78 (2001) 1276.

[43] T. Devillers, D. M. G. Leite, J. H. Dias da Silva, A. Bonanni, Functional Mn–Mg$_k$ cation complexes in GaN featured by Raman spectroscopy, Appl. Phys. Lett. 103 (2013) 211909.

[44] J. Suffczynski, A. Grois, W. Pacuski, A. Golnik, J. A. Gaj, A. Navarro-Quezada, B. Faina, T. Devillers, A. Bonanni, Effects of s,p-d and s-p exchange interactions probed by exciton magnetospectroscopy in (Ga,Mn)N, Phys. Rev. B 83 (2001) 094421.

[45] Light Scattering in Solid I, edited by M. Cardona (Springer, New York, 1983).

[46] R. M. Martin and T. C. Damen, Breakdown of selection rules in resonance Raman scattering, Phys. Rev. Lett. 26 (1971) 86.

[47] M. Zajac, R. Kucharski, K. Grabianska, A. Gwardys-Bak, A. Puchalski, D. Wasik, E. Litwin-Staszewska, R. Piotrzkowski, J.Z. Domagala, M. Bockowski, Basic ammonothermal growth of Gallium Nitride - state of the art, challenges, perspectives, Prog. Cryst. Growth Char. Mater. 64 (2018) 63.

[48] A. Bonanni, M. Sawicki, T. Devillers, W. Stefanowicz, B. Faina, T. Li, T.E. Winkler, D. Sztenkiel, A. Navarro-Quezada, M. Rovezzi, R. Jakieła, A. Grois, M. Wegscheider, W. Jantsch, J. Suffczynski, F. D'Acapito, A. Meingast, G. Kothleitner, T. Dietl, Experimental probing of exchange interactions between localized spins in the dilute magnetic insulator (Ga,Mn)N, Phys. Rev. B 84 (2011) 035206.

[49] T. Yamamoto, H. Sazawa, N. Nishikawa, M. Kiuchi, T. Ide, M. Shimizu, T. Inoue, M. Hata, Reduction in buffer leakage current with Mn-doped GaN buffer layer grown by metal organic chemical vapor deposition, Jpn. J. Appl. Phys. 52 (2013) 08JN12.

[50] R. Adhikari, W. Stefanowicz, B. Faina, G. Capuzzo, M. Sawicki, T. Dietl, A. Bonanni, Upper bound for the s-d exchange integral in n-(Ga,Mn)N:Si from magnetotransport studies, Phys. Rev. B 91 (2015) 205204.

[51] L. L. Guo, Y. H. Zhang, W. Z. Shen, Temperature dependence of Raman scattering in GaMnN, Appl. Phys. Lett. 89 (2006) 161920.

[52] J. H. D. da Silva, D. M. G. Leite, A. Tabata, A. A. Cavalheiro, Structural and vibrational analysis of nanocrystalline $Ga_{1-x}Mn_xN$ films deposited by reactive magnetron sputtering, J. Appl. Phys. 102 (2007) 063526.

[53] M. Sawicki, Magnetic properties of (Ga,Mn)As, J. Magn. Magn. Mat. 300 (2006) 1.

[54] A. Twardowski, H. J. M. Swagten, W. J. M. de Jonge, M. Demianiuk, Magnetic behavior of the diluted magnetic semiconductor $Zn_{1-x}Mn_xSe$, Phys. Rev. B 36 (1987) 7013.